\documentclass[a4paper]{jpconf}
\usepackage{graphicx}
\begin{document}
\title{Quantum-gravitational corrections to the power spectrum for a closed universe}

\author{Tatevik Vardanyan, Claus Kiefer}

\address{University of Cologne, Faculty of Mathematics and Natural Sciences, Institute for Theoretical Physics, Cologne, Germany}

\ead{tatevik@thp.uni-koeln.de, kiefer@thp.uni-koeln.de}

\begin{abstract}
We study the quantum-gravitational corrections to the power spectrum of a gauge-
invariant inflationary scalar perturbations in a closed model of a universe. We consider canonical
quantum gravity as an approach to quantizing gravity. This leads to the Wheeler-DeWitt equation,
which has been studied by applying a semiclassical Born–Oppenheimer type of approximation. At
the corresponding orders of approximation, we recover both the uncorrected and quantum-gravitationally corrected Schr\"{o}dinger equations for the perturbation modes from which we
calculate the quantum-gravitational corrections to the power spectrum in the slow-roll regime.
The results are compared to the power spectra for the flat model of the universe.
\end{abstract}

\section{Introduction}

Quantum-gravitational effects are expected to be strong at very high energies. Such energies have been present in the very early universe; therefore one hopes to find quantum-gravitational effects related to the anisotropies of the CMB. One assumes that these anisotropies originate from quantum fluctuations of the inflationary field and the metric. 

To study the quantum-gravitational corrections to the power spectrum of inflationary perturbations, we will consider canonical quantum gravity as a conservative approach to quantize gravity. This starts with the Hamiltonian formulation of general relativity (GR) and leads to the Wheeler-DeWitt equation \cite{book:Kiefer}. Applying a semiclassical approximation, we obtain a Schr\"{o}dinger equation for the perturbation modes including quantum-gravitational correction terms, from which we calculate the power spectrum.

We consider a closed model of the universe because of the following arguments. The recent Planck observation \cite{paper:Planck2018} is consistent with a flat $\Lambda$CDM model; however, the data doesn't rule out the possibility that our universe might have slight positive curvature.  
In \cite{paper:Valentino}, the authors argue that from Planck data, the positive curvature is preferred for more than $99\%$. They claim that the positive curvature explains the anomalous lensing amplitude as well as it removes the tension within the Planck data set for the values of cosmological parameters obtained at different angular scales. In addition, a closed model might explain the observed low amplitude of quadrupole and octopole modes \cite{paper:Planck2018}, which remains unexplained in the standard $\Lambda{\rm CDM}$ model of cosmology.

Furthermore, even though inflation would flatten any initial spatial curvature (assuming it is present before the inflation), the presence of initial spatial curvature could have a significant imprint on the power spectrum, the study of which is one of the main goals of this paper.  

Finally, there are debates regarding infinities in science from mathematical and philosophical perspective \cite{paper:EMN}. Thus, considering a closed model of a universe, for which the space is finite, serves to a purpose of tackling these subtle issues as well.

In what follows, in section 2 we present the general features of canonical quantum gravity
and apply this quantization scheme to the inflationary universe. We derive the Wheeler-DeWitt equation for the inflationary universe and in section 3 we apply a semiclassical approximation
scheme and obtain the uncorrected and the quantum-gravitationally corrected Schr\"{o}dinger
equations. Finally, in section 4, we present the derived uncorrected and quantum-gravitationally
corrected power spectra. The results are summarized in section 5. Technical details can be found in \cite{paper:CT}, \cite{thesis:Tatevik}.

\section{Derivation of the Wheeler-DeWitt equation for an inflationary closed universe}
Let us briefly present the canonical quantization procedure of gravity. The first step is the construction of the Hamiltonian formulation for GR which starts with the foliation of spacetime into three-dimensional spacelike Cauchy hypersurfaces parameterized by a time function $t$ \cite{book:Kiefer}.
The spacetime foliation leads to the following line element
\begin{equation}
    d s^2=-(N^2-N_i N^i)d t^2+2N_i d x^i d t +h_{ij} d x^i d x^j,
    \label{ADM}
\end{equation}
where $N$ is the lapse function, $N_i$ is the shift vector, and $h_{ij}$ is the three-metric.

Rewriting the action using (\ref{ADM}) and taking variations of the action with respect to the lapse function and shift vector leads to the Hamiltonian and momentum constraints.

According to the procedure for the quantization of constrained systems, a classical constraint converts to quantum operators and turns into a restriction on the wave functional. 
The quantization of the Hamiltonian constraint leads to the Wheeler-DeWitt equation and, respectively, the momentum constraint to the quantum diffeomorphism constraint. 

\subsection{Hamiltonian formulation}
Let us now apply the above procedure to the inflationary Friedmann universe.
We impose the FLRW line element 
\begin{equation}
 d s^{2}=-N^{2}(t)d t^{2}+a^{2}(t)d\Omega^{2}_{3},
\end{equation}
where $a(t)$ is the scale factor, $d\Omega_3^2$ denotes the metric of the unit three-sphere, and due to the symmetry only the lapse function $N(t)$ enters in the line element. 

Including perturbations to the metric and the scalar field $\phi$ with a quadratic potential $\mathcal{V}(\phi)=\frac{1}{2}m^2\phi^2$, it can be shown that the total Hamiltonian can be written as a sum of the background and perturbation parts \cite{paper:Hawking},
\begin{equation}
H=N\Big[H_{|0}+\sum_{n}{H_{|2}^n} \Big].
\label{LH0}
\end{equation} 
The the background Hamiltonian reads
\begin{equation}
H_{|0}=N\frac12 e^{-3\alpha}\Big[-\frac{1}{m_{\rm P}^2 }{p_{\alpha}}^2+{p_{\phi}}^2+2e^{6\alpha}\mathcal{V}(\phi)-m_{\rm P}\mathcal{K}e^{4\alpha }\Big],
\label{H0}
\end{equation}
where
\begin{equation}
p_{\alpha}=-m_{\rm P}^2e^{3\alpha}\frac{\dot{\alpha}}{N}\ \ \ \textnormal{and} \ \ \ p_{\phi}=e^{3\alpha}\frac{\dot{\phi}}{N}, \ \ m_{\rm P}:=\sqrt{{3\pi}/{2 G}},\ \ \alpha=\ln{a/a_0}.
  \end{equation} 
The curvature parameter $\mathcal{K}$ takes the value $\mathcal{K}=1$ for the closed model. 

The perturbation Hamiltonian, denoted as ${H_{|2}^n}$, is composed of scalar, vector and tensor parts. Since we are interested in scalar perturbations the vector and tensor parts will be ignored. 
In addition, the first-order Hamiltonian constraints
\begin{equation}
 H^n_{\ |1}=0, \ \ \ {^S{H^n_{\ \_1}}}=0,\ \ \ {^V{H^n_{\,\_1}}}=0
\end{equation}
are the linearized energy constraint and the scalar and vector parts of the linearized momentum constraints correspondingly. 

The perturbation Hamiltonian in (\ref{LH0}) is not gauge-invariant. The gauge-invariant form for the scalar part of the perturbation Hamiltonian (which will be denoted by $^S{H_{|2}^n}$) was obtained in \cite{paper:Langlois}. To bring this gauge-invariant Hamiltonian into a more compact form we perform a canonical transformation which leads to the following expression: 
 \begin{equation}
{^s{H_{|2}^n}}=\frac12\Big[{\tilde{p}_n}^2+\omega_n^2(\eta){\tilde{v}_n}^2  \Big],
\end{equation}
where $\eta$ is the conformal time, $\tilde{v}_n$ is the gauge-invariant variable and $ \tilde{p}_n$ is its conjugate momentum.  
The explicit expression of the function $\omega_n^2(\eta)$ can be found in \cite{paper:CT}.

In the large-wavelength limit the well-known result for the perturbation Hamiltonian in the flat model \cite{paper:Martin} is recovered.

\subsection{Quantization}
We consider the slow-roll regime for which the scalar field satisfies the following conditions
\begin{equation}
 \dot{\phi}\ll\mathcal{V}(\phi), \quad  \ddot{\phi}\ll3 H\phi.
\end{equation}
Introducing the slow-roll parameters defined by $\varepsilon=1-\frac{\mathcal{H}^{\prime}}{\mathcal{H}^{2}}$ and $ \delta=\varepsilon-\frac{\varepsilon^{\prime}}{2\mathcal{H}\varepsilon}$,\footnote{A dimensionless Hubble parameter $\mathcal{H}$ is defined by $\mathcal{H}=a^\prime/a$. The derivative with respect to $\eta$ is denoted by a prime.} the Wheeler-DeWitt equation for each mode takes the following form
\begin{equation}
\frac12\Bigg\{ e^{-2\alpha}\Bigg[ \frac{1}{m_{\rm P}^2}\frac{\partial^2}{\partial \alpha^2}+m_{\rm P}^2\Bigg( e^{6\alpha}{H}^2\Big(1-\frac{\varepsilon}{3}\Big)-\frac{\mathcal{K}}{3}e^{4\alpha}\Bigg)\Bigg]-\frac{\partial^2}{\partial {\tilde{v}_n}^2 }+\omega_n^2(\eta){\tilde{v}_n}^2 \Bigg\}\Psi_n\big(\alpha, \tilde{\phi},{\tilde{v}_n}\big) =0.
\label{WD}
 \end{equation}
 where we have introduced a dimensionless variable $\tilde{\phi}=m_{\rm P}^{-1}\phi$.
To obtain the above equation we have made a product ansatz for the full wave functional assuming that the perturbation modes do not interact with each other. 

\section{Semiclassical approximation}
We apply a semiclassical Born-Oppenheimer type of approximation which allows us to obtain the Schr\"{o}dinger equation for the perturbations propagating on the classical background and the quantum-gravitational corrections to it following the procedure developed in \cite{paper:Manuel}, \cite{paper:David}. 

We make a WKB-like ansatz
\begin{equation}
   \Psi_n\big(\alpha,\tilde{\phi},{\tilde{v}_n}\big)=e^{i S(\alpha, \tilde{\phi},\tilde{v}_n)}.
   \label{wkba}
\end{equation}
The Born-Oppenheimer approximation is implemented by expanding $S(\alpha, \phi,\tilde{v}_n)$ in terms of powers of Planck mass
\begin{equation}
 S(\alpha,\tilde{v}_n)=m_{\rm P}^2S_0+m_{\rm P}^0S_1+m_{\rm P}^{-2}S_2+ \ldots .  
\end{equation}
Afterwards, we insert (\ref{wkba}) into the Wheeler-DeWitt equation (\ref{WD}), collect the corresponding terms of the powers of $m_{\rm P}$ and set them equal to zero.

At order $m_{\rm P}^2$, the Hamilton--Jacobi equation for the classical background is obtained which is equivalent to the Friedmann equation. At the next order, $m_{\rm P}^0$, the obtained equation can be rewritten as a Schr\"{o}dinger equation for a wave functions $\psi_n^{(0)}$, i.e.
\begin{equation}
     i\frac{\partial}{\partial \eta}\psi_n^{(0)}={^s{H_{|2}^n}}\psi_n^{(0)},
     \label{Se}
\end{equation}
where the conformal time parameter $\eta$ has been introduced as
\begin{equation}
    \frac{\partial}{\partial \eta}:=-e^{-2\alpha}\frac{{\partial}S_0}{\partial \alpha}\frac{{\partial}}{\partial{\alpha}}= e^{\alpha}\sqrt{ H^2\Big(1-\frac{\varepsilon}{3}\Big)-\frac{\mathcal{K}}{3}e^{-2\alpha}}\frac{{\partial}}{\partial{\alpha}}. 
\end{equation}
At the order of $m_{\rm P}^{-2}$, we obtain the quantum-gravitational corrections to the Schr\"{o}dinger equation following the procedure in \cite{paper:Manuel},
\begin{equation}
  i\frac{\partial}{\partial \eta} \psi_n^{(1)}= {^s{H_{|2}^n}} \psi_n^{(1)}-\frac{\psi_n^{(1)}}{2m_{\rm P}^2 \psi_n^{(0)}}\left[i\frac{\partial}{\partial \eta}\left(\frac{{^s{H_{|2}^n}}}{{V}}\right)\psi_n^{(0)}+\frac{ 1}{{{V}}} \big({^s{H_{|2}^n}}\big)^2 \psi_n^{(0)}\right],
  \label{cSe}
\end{equation}
where $V\left(a,\tilde{\phi}\right):=\frac{2 a^4}{m_{\rm P}^2} \mathcal{V}\left(\tilde{\phi}\right)$.
Note that on the right-hand side, there is a unitarity violating term. In this paper, following the arguments in \cite{paper:David} we will drop the non-unitarity term because of its unphysical behavior. However, the treatment of this term will be carried on in an upcoming work with L. Chataignier \cite{paper:LT} following the method developed in \cite{paper:CK2021}, where the
authors argue that this non-unitary term can be absorbed by using a suitable definition of the inner product. 

To solve the Schr\"{o}dinger equation (\ref{Se}) we make a Gaussian ansatz, 
\begin{equation}   
  \psi_n^{(0)}(\eta,\tilde{v}_n)=A_n^{(0)}(\eta)e^{-\frac{1}{2}\Omega_n^{(0)}(\eta)\tilde{v}_n^2}.
\end{equation}
Inserting this into the Schr\"{o}dinger equation we get
\begin{equation}   
{ \Omega_n^{(0)}}^{\prime}(\eta)=-i\Big[ {\big(\Omega_n^{(0)}(\eta)\big)}^2-\omega_n^2\Big].
\label{Omega0}
\end{equation}
Similarly, using the following Gaussian ansatz for the corrected case (\ref{cSe}), 
\begin{equation}
 \psi_n^{(1)}(\eta,\tilde{v}_n)=\Big(A_n^{(0)}(\eta)+\frac{1}{m_{\rm P}^2}A_n^{(1)}(\eta)\Big)\textnormal{exp}{\Big[-\frac12\Big(\Omega_n^{(0)}(\eta)+\frac{1}{m_{\rm P}^2}\Omega_n^{(1)}(\eta)\Big)\tilde{v}_n^2\Big]},
\end{equation}
leads to the equation 
\begin{equation}
{\Omega_n^{(1)}}^{\prime}(\eta)
   = 2 \Omega_n^{(0)}(\eta)\Big[\Omega_n^{(1)}(\eta)
  -\frac{3}{4{{V}}}\Big(\big({\Omega_n^{(0)}\big)}^2(\eta)-\omega_n^2(\eta)\Big)\Big].
  \label{Omega1}
\end{equation}

\section{Derivation of the power spectra}
\subsection{Uncorrected power spectrum}
Starting with the two-point correlation function one obtains the following expression for the power spectrum:
\begin{equation}
 \mathcal{P}_{\tilde{v}}= \frac{n(n^2-\mathcal{K})}{2 \pi^2}\frac{1}{2\mathcal{R}e \Omega_n^{(0)}}.
 \label{PowerS}
\end{equation}  
Solving (\ref{Omega0}) and substituting the real part of $\Omega_n^{(0)}(\eta)$ into (\ref{PowerS}) leads to 
\begin{equation}
 \mathcal{P}_{\tilde{v}}
  =\frac{n^2-{\mathcal{K}}}{4 \pi^2 f_\varepsilon^3(n)\xi^2}\Big[\tilde{C}_{\varepsilon,\delta}-2(\gamma-3\lambda_{\varepsilon,\delta}(n)/2)\ln\big(\xi f_\varepsilon(n)\big)\Big], \ \ \textnormal{with} \ \ \xi=\frac{n}{a H}.
  \label{PS}
\end{equation}
For the explicit expressions of the functions $\lambda_{\varepsilon,\delta}(n)$, 
$f_\varepsilon(n)$ and $\tilde{C}_{\varepsilon,\delta}$ see \cite{paper:CT}.

To relate the power spectrum of scalar perturbations at the end of inflation to the temperature anisotropies of the CMB, one needs the power spectrum of curvature perturbations. Hence, we introduce a gauge-invariant variable $\zeta_{\rm BST}$\footnote{For the flat model another variable $\xi$ is used which cannot be used here (see the details in \cite{paper:CT}).} which can be related to the variable $\tilde{v}_n$. In the slow-roll regime the relation is given by  
\begin{equation}
 \zeta_{{\rm BST},n}=-\frac{1}{a}\frac{1}{\sqrt{2\left(\varepsilon+\frac{\mathcal{K}}{\mathcal{H}^2}\right)}M_{\rm P}}\Bigg[1+\frac{\mathcal{K}}{n^2-4\mathcal{K}}\left(\varepsilon+\frac{\mathcal{K}}{\mathcal{H}^2}\right)\Bigg]^{1/2}\tilde{v}_n,
 \end{equation}
where we have introduced a redefined Planck mass $M_{\rm P}:=1/\sqrt{8\pi G}$.
The power spectrum of $\zeta_{\rm BST}$ reads
\begin{equation}
    \mathcal{P}_{\zeta_{\rm BST}}= \frac{1}{2a^2\left(\varepsilon+\frac{\mathcal{K}}{\mathcal{H}^2}\right)M_{\rm P}^2}\Bigg[1+\frac{\mathcal{K}}{n^2-4\mathcal{K}}\left(\varepsilon+\frac{\mathcal{K}}{\mathcal{H}^2}\right)\Bigg] \mathcal{P}_{\tilde{v}_n}.
    \label{PzBST}
\end{equation}
In the large-wavelength limit, we recover the power spectrum for the flat model \cite{paper:Manuel}. 

\begin{figure}
\centering
\includegraphics[width=0.5\textwidth]{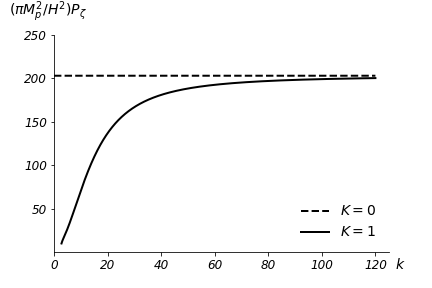}
\caption{Plot of the power spectra of curvature perturbation at the point of horizon reentry, i.e. $\xi=1$s . The flat case is presented with the dashed line and closed one with the thick line. The slow-roll parameters are set to $\varepsilon=0.005$ and $\delta=-0.006$. Figure adapted from \cite{paper:CT}.}\label{PlotPz}
\end{figure}

In Fig.~\ref{PlotPz}, the power spectra of curvature perturbations are plotted as a function of the comoving wavelength $k$ for the flat and the closed models. One can see that there is a significant suppression of power at large scales for the closed case compared to the flat case. Moreover, due to the finiteness of the universe, there is a cutoff at a small $k$.

Finally, let us point out that due to the explicit scale dependence of the power spectrum (\ref{PzBST}) we cannot apply a power-law approximation at large scales where the scale dependence is the most prominent. The power-law approximation is based on the assumption of weak scale dependence.  

\subsection{Corrected power spectrum}
The quantum-gravitationally corrected power spectrum takes the form  \cite{thesis:Tatevik}, \cite{dis:Manuel}
\begin{eqnarray}
    \mathcal{P}_{\tilde{v}}^{(1)}(n)=&\frac{n(n^2-\mathcal{K})}{4 \pi^2}\Bigg(\mathcal{R}e \Omega_n^{(0)}+\frac{1}{m_{\rm P}^2}\mathcal{R}e \Omega_n^{(1)}\Bigg)^{-1}.
\end{eqnarray}
Solving the equation (\ref{Omega1}) and substituting $\mathcal{R}e \Omega_n^{(1)}$ into the above relation, the power spectrum takes the form \cite{thesis:Tatevik}
\begin{eqnarray}
  \mathcal{P}_{\tilde{v}}^{(1)}(n)= \mathcal{P}_{\tilde{v}}^{(0)}(n)\Big(1+ \Delta_{\tilde{v}}\Big),   \  \textnormal{where} \ \Delta_{\tilde{v}}=-\frac{H^2}{m_{\rm P}^2}\frac{\Big[ \tilde{C}_{\varepsilon,\delta}-2(\gamma-3\lambda/2)\ln{\big(\xi f(n,\varepsilon)\big)}\Big]}{n^3 f^3(n,\varepsilon)\xi^2}\mathcal{R}e \tilde{\Omega}_n^{(1)}(\xi).
\end{eqnarray}
The corrected power spectrum for the flat case \cite{paper:Manuel} is recovered by considering the large wavelength limit.

It was estimated in \cite{paper:Manuel} that the corrections $|\Delta_v|\sim 10^{-10}$. Since the corrections are so small and are within the experimental error bars, unfortunately, we don't expect to be able to observe quantum-gravitational corrections to the power spectrum in the near future.

\section{Conclusion}

We have calculated the quantum-gravitational corrections to the power spectrum of gauge-invariant inflationary scalar perturbations in a closed model for the universe. We have derived the Wheeler-DeWitt equation for the inflationary universe and applied a semiclassical Born–Oppenheimer type of approximation. At the corresponding orders of approximation, we obtained the uncorrected and quantum-gravitationally corrected Schr\"{o}dinger equations from which we have derived the uncorrected and corrected power spectra. 

We have also calculated the power spectrum of curvature perturbations which exhibits suppression of power at large scales compared to the flat model; hence, it may explain the observed lack of power for large scales. 

Regarding the quantum-gravitationally corrected power spectrum, the correction leads to an additional suppression of power at large scales. 
Unfortunately, due to Cosmic Variance there is significant statistical uncertainty at the largest scales where this contribution is expected. Therefore, we do not expect to confirm these by observations in the near future. We emphasize, though, that this is a concrete quantitative prediction from quantum gravity.

\section{Acknowledgements}
We would like to thank Leonardo Chataignier for fruitful discussions. T. V. thanks the organizers of DICE2022 for providing the opportunity to present a talk.

\end{document}